\documentclass[runningheads]{llncs}
\usepackage[T1]{fontenc}
\usepackage{times}
\usepackage{amsmath}
\usepackage{amssymb}
\usepackage{multicol,multirow}
\usepackage[tableposition=top]{caption}
\usepackage{cite}
\usepackage{booktabs}
\usepackage{bigstrut}
\usepackage{floatrow}
\usepackage{rotating}
\usepackage{adjustbox}
\usepackage{amsmath}

\usepackage{graphicx}
\begin{document}
\title{EUIS-Net: A Convolutional Neural Network for Efficient Ultrasound Image Segmentation}
%
%\titlerunning{Abbreviated paper title}
% If the paper title is too long for the running head, you can set
% an abbreviated paper title here
%
\author{Shahzaib Iqbal \inst{1}\orcidID{0000-1111-2222-3333} \and
Hasnat Ahmed\inst{1} \and
Muhammad Sharif\inst{1} \and
Madiha Hena\inst{2} \and
Tariq M. Khan\inst{3} \and
Imran Razzak\inst{3}}
\authorrunning{S. Iqbal et al.}
% First names are abbreviated in the running head.
% If there are more than two authors, 'et al.' is used.
%
\institute{Department of Electrical Engineering, Abasyn University Islamabad Campus(AUIC), Islamabad, Pakistan (\email{shahzeb.iqbal@abasynisb.edu.pk}) \and
Department of Computing, Abasyn University Islamabad Campus(AUIC), Islamabad, Pakistan (\email{madiha.hena@abasynisb.edu.pk})\\
% \url{http://www.springer.com/gp/computer-science/lncs} 
\and
School of Computer Science \& Engineering, UNSW, Sydney, Australia\\
\email{\{tariq.khan,imran.razzak\}@unsw.edu.au}}
\maketitle              % typeset the header of the contribution
\begin{abstract}
Segmenting ultrasound images is critical for various medical applications, but it offers significant challenges due to ultrasound images' inherent noise and unpredictability. To address these challenges, we proposed EUIS-Net, a CNN network designed to segment ultrasound images efficiently and precisely. The proposed EUIS-Net utilises four encoder-decoder blocks, resulting in a notable decrease in computational complexity while achieving excellent performance. The proposed EUIS-Net integrates both channel and spatial attention mechanisms into the bottleneck to improve feature representation and collect significant contextual information. In addition, EUIS-Net incorporates a region-aware attention module in skip connections, which enhances the ability to concentrate on the region of the injury. To enable thorough information exchange across various network blocks, skip connection aggregation is employed from the network's lowermost to the uppermost block. Comprehensive evaluations are conducted on two publicly available ultrasound image segmentation datasets. The proposed EUIS-Net achieved mean IoU and dice scores of 78. 12\%, 85. 42\% and 84. 73\%, 89. 01\% in the BUSI and DDTI datasets, respectively. The findings of our study showcase the substantial capabilities of EUIS-Net for immediate use in clinical settings and its versatility in various ultrasound imaging tasks.

\end{abstract}
\section{Introduction}
Breast cancer remains a prevalent disease and a leading cause of mortality among women worldwide, as highlighted by the World Health Organisation (WHO) \cite{WHO2023}. In 2020, breast cancer alone accounted for 2.3 million new cases and 685,000 deaths worldwide \cite{islami2019national}. Projections indicate that by 2040, these numbers could reach more than 3 million new cases and approximately 1 million deaths annually \cite{world2023global}. Various imaging modalities, including mammography, computed tomography (CT), magnetic resonance imaging (MRI), and breast ultrasound (BUS), play crucial roles in breast cancer detection. Mammography, the standard diagnostic tool, has limitations due to its use of ionising radiation, making it unsuitable for pregnant women \cite{skaane2019digital}. In contrast, BUS imaging is both cost-effective and radiation-free, making it an invaluable screening tool \cite{carlino2019ultrasound}. It provides detailed information on the characteristics of breast tissue and the presence of malignant tissues \cite{carlino2019ultrasound}.

Segmentation of breast lesions from ultrasound images has been extensively investigated using various algorithms. These methods can be broadly categorised into early techniques and machine learning approaches \cite{khan2020machine}. Early techniques include methods such as graph-based models, region-growing models, and deformable models. Machine learning approaches encompass traditional hand-made methods and more recent deep learning (DL) techniques \cite{iqbal2022recent,khan2023retinal,khan2023feature,iqbal2023ldmres}.

Graph-based models optimise energy using frameworks like Markov random fields or graph cuts. Chiang et al. \cite{chiang2010cell} employed a pre-trained Probabilistic Boosting Tree (PBT) classifier to compute the data term within graph-cut energy, while Xian et al. \cite{xian2015fully} integrated information from both frequency and spatial domains to formulate the energy function. However, these models may struggle to capture intricate semantic features and delineate faint boundaries in low-contrast ultrasound images, leading to inaccuracies.

Deformable models start with an initial representation and iteratively deform toward object boundaries, guided by internal and external energy factors. For example, Madabhushi et al. \cite{madabhushi2003combining} used boundary points and balloon forces to define the external energy in their deformable model. Chang et al. \cite{madabhushi2003combining} reduced speckle noise in ultrasound images using a stick filter and applied a discrete 3D active contour model to accurately segment regions of breast lesions by deforming the model.

Region-growing methods initiate segmentation by starting from manually or automatically selected seed points and gradually expanding to delineate target region boundaries based on predefined growth criteria. Shan et al. \cite{shan2012completely} used this approach to segment breast cancer, incorporating criteria such as contour smoothness and region similarity to enhance the segmentation process.

Machine learning-based approaches have traditionally relied on hand-crafted features to train classifiers for segmentation tasks \cite{khan2024esdmr, naveed2024ra, khan2022leveraging}. In contrast, recent advances in deep learning, particularly convolutional neural networks (CNN), have shown promise in automatically learning high-level features from data \cite{javed2024advancing, matloob2024lmbis, khan2024lmbf}. For example, models such as ESTAN \cite{shareef2022estan}, RCA-IUnet \cite{punn2022rca}, CTG-Net \cite{yang2022ctg}, and BGM-Net \cite{wu2021bgm} have been developed specifically for breast tumour segmentation in ultrasound images. These models incorporate innovations such as attention mechanisms, multiscale features, and boundary-guided networks to improve segmentation accuracy.

% This task is especially challenging in breast ultrasound (BUS) imaging due to the variable shapes of tumors, indistinct contours, low contrast, and inherent noise \cite{zhou2021multi}. Recent advancements in CAD technology have led to investigations into its efficacy for breast cancer detection and segmentation using BUS images \cite{drukker2008breast,yap2010processed}. Despite these challenges, CAD systems have shown significant progress in accurately detecting and segmenting lesions.

\begin{figure*}[h]
    \centering
    \includegraphics[width=0.85\textwidth]{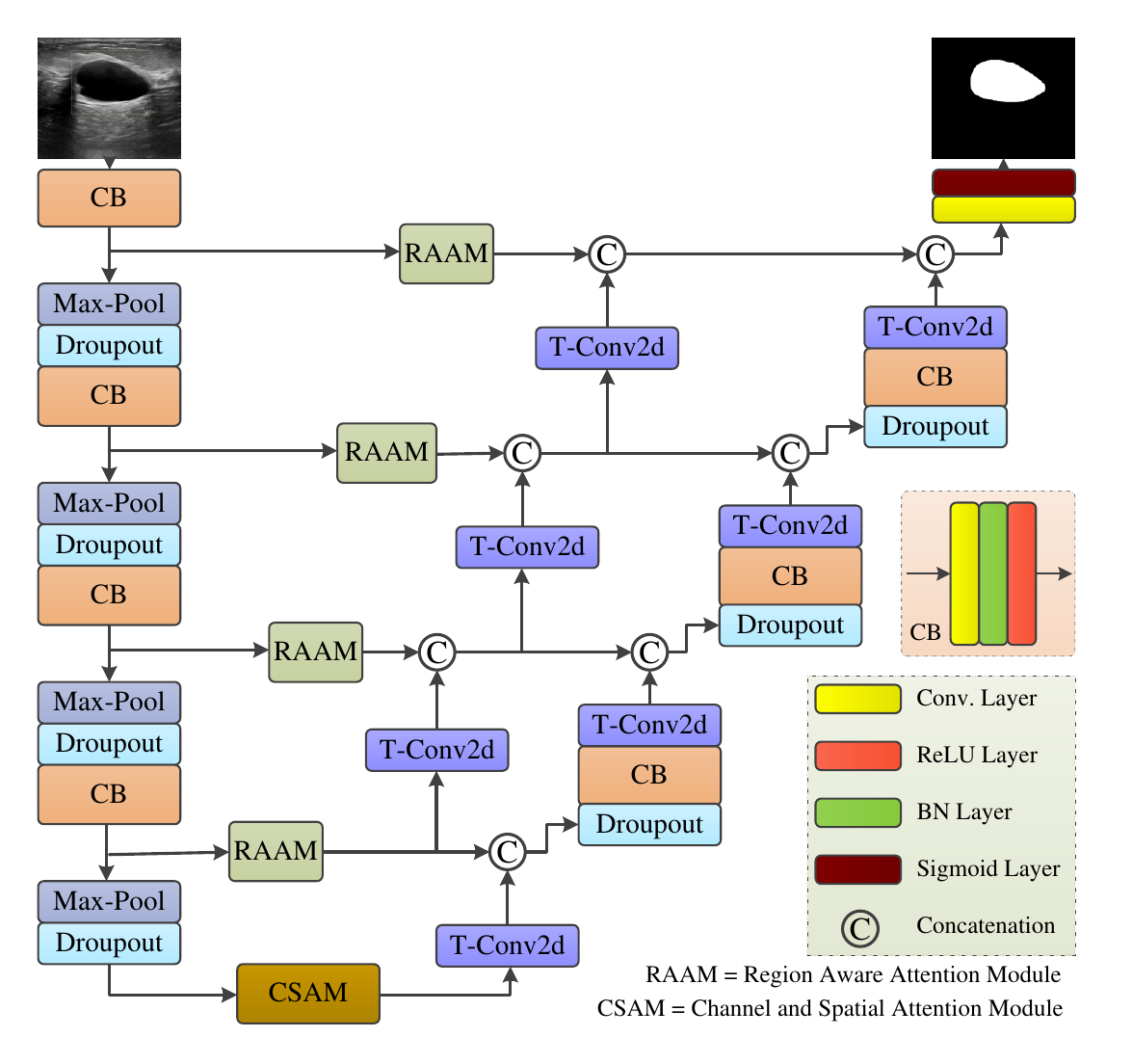}
    \caption{Schematic of the proposed method. }
    \label{fig:model}
\end{figure*}

\begin{figure*}[h]
    \centering
    \includegraphics[width=\textwidth]{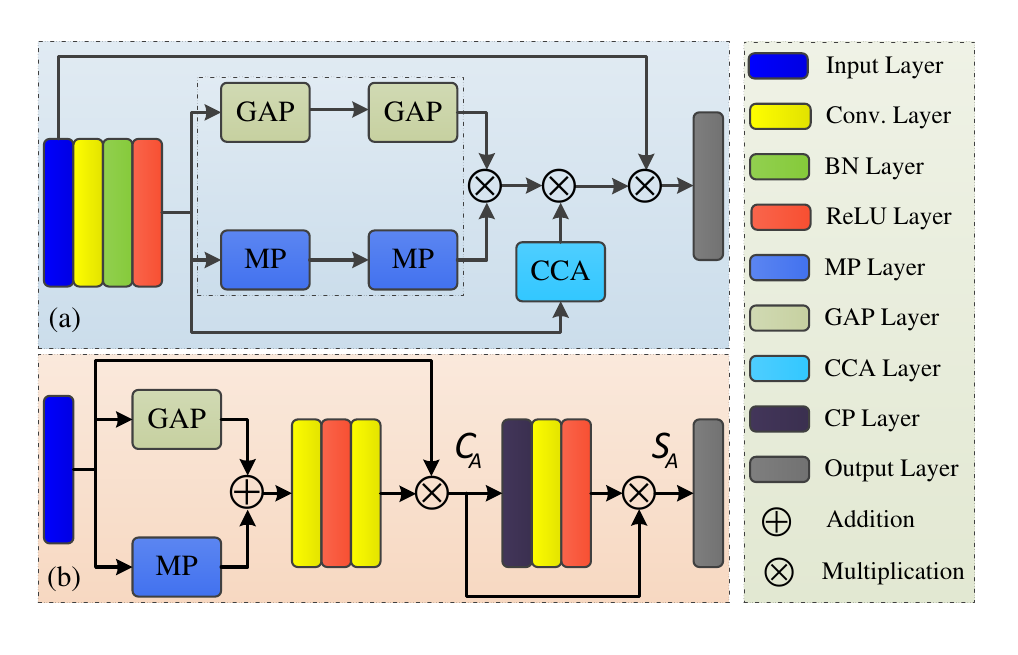}
    \caption{Schematic of the attention modules: (a) region aware attention module, (b) channel and spatial attention module. }
    \label{fig:modules}
\end{figure*}
% In contrast, segmentation methods based on deep learning have shown significant performance gains in detecting breast lesions \cite{amin2020distinctive,amin2022breast,shah2020facial,sharif2020framework,umer2022integrated,umer2022multi}. These methods utilize various deep learning techniques for segmenting breast cancer from BUS images, leveraging automated feature learning with convolutional layers.

Despite progress, current methods still face significant challenges. Region-growing methods are hindered by noise and variable tissue properties, while graph models often fail to capture complex tumor features in low-contrast images. Deformable models are highly sensitive to initialisation, leading to suboptimal results depending on the starting conditions. Handcrafted feature approaches may not fully exploit the extensive information present in the data. Although deep learning methods are promising, they require further refinement to enhance segmentation accuracy.

To address these challenges, we proposed EUIS-Net, a CNN network optimized for efficient and precise ultrasound image segmentation. EUIS-Net employs four encoder-decoder blocks, significantly reducing computational complexity while maintaining exceptional performance. The network integrates both channel and spatial attention mechanisms in the bottleneck to enhance feature representation and capture critical contextual information. In addition, it incorporates a region-aware attention module (RAAM) into the skip connections, improving the focus on the areas of the injury. To ensure comprehensive information exchange across different network blocks, skip connection aggregation is used from the lowest to the highest block.

The remainder of this manuscript is structured as follows.
Section \ref{method} details the architecture of the proposed EUIS-Net model. Section \ref{experimentalResults} presents the experimental results and discussions. Finally, Section \ref{conclusions} concludes the paper and outlines potential future research directions.

\section{Proposed Method}\label{method}
In this section, the proposed EUIS-Net is discussed briefly. The proposed EUIS-Net starts with a feature extraction block that employs a convolutional block ($CB$) followed by a max-pooling operation and a dropout layer. The encoder of the proposed EUIS-Net uses four feature extraction blocks. Once feature information is extracted on multiple spatial scales, these features are further refined by channel and spatial attention modules at the bottleneck layer. The proposed EUIS-Net decoder uses the same convolutional blocks to reconstruct the feature maps in their actual spatial dimension. The proposed method also uses region-aware attention modules \cite{naveed2024ra} in the skip connection to refine and control the feature information flow between the encoder and decoder blocks.

Let $In$ be the RGB image given as input to the proposed method. The output of the first encoder block ($B_{E}^{1}$) is computed by applying a convolutional block ($\texttt{CB}$) on the input as given in (Eq. \ref{Eq:1}). 

\begin{equation}
    B_{E}^{1}=\texttt{CB}(In)
    \label{Eq:1}
\end{equation}

where $CB$ is the convolutional block that uses the standard convolutional operation followed by a batch normalisation layer and activation function computed in (Eq.\ref{CB}).

\begin{equation}
    \texttt{CB}=\Re \left ( B_{n}\left ( f^{3\times 3}\left ( In \right ) \right ) \right )
    \label{CB}
\end{equation}
where ($f^{3 \times 3}$) is the standard convolutional operation with a kernel size of ($3 \times 3$), $B_{n}$ is the batch normalization and $\Re$ is the ReLU activation layer. The output of the $2^{nd}, 3^{rd}$ and $4^{th}$ encoder blocks are computed as (Eqs. \ref{Eq:2}--\ref{Eq:4}).

\begin{equation}
    B_{E}^{2}=\texttt{CB}\left ( D_{r}\left ( M_{p}\left ( B_{E}^{1} \right ) \right ) \right )
    \label{Eq:2}
\end{equation}

\begin{equation}
    B_{E}^{3}=\texttt{CB}\left ( D_{r}\left ( M_{p}\left ( B_{E}^{2} \right ) \right ) \right )
    \label{Eq:3}
\end{equation}

\begin{equation}
 B_{E}^{4}=\texttt{CB}\left ( D_{r}\left ( M_{p}\left ( B_{E}^{3} \right ) \right ) \right )
    \label{Eq:4}
\end{equation}

where ($D_{r}$) is the dropout layer and ($M_{p}$) is the max-pooling operation. Once the feature information is extracted, it is refined by the channel and spatial attention module (CSAM). The output of the $1^{st}$ decoder block ($B_{D}^{1}$) is calculated employing a max-pooling operation followed by dropout, $CSAM$, and transposed convolution (for upsampling) on the $4^{th}$ encoder block ($B_{E}^{4}$) and then concatenating it with the output of the region-aware attention module ($RAAM$) applied on the encoder block $4^{th}$ ($B_{E}^{4}$) as computed in (Eq. \ref{Eq:D1}). 

\begin{equation}
B_{D}^{1}=\left [ f_{t}^{3\times 3}\left ( \texttt{CSAM}\left ( D_{r}\left ( M_{p}\left ( B_{E}^{4} \right ) \right ) \right ) \right ) \right ] \copyright \left [ \texttt{RAAM}\left ( B_{E}^{4} \right ) \right ]
    \label{Eq:D1}
\end{equation}
where  $f_{t}^{3\times 3}$ is the transposed convolution operation with a kernel size of ($3\times 3$) and $\copyright$ is the concatenation operation. The output of the $2^{nd}, 3^{rd}$ and $4^{th}$ decoder blocks are computed as (Eqs. \ref{Eq:D2}--\ref{Eq:D4}).

\begin{equation}
    B_{D}^{2}= \left [ f_{t}^{3\times 3}\left ( \texttt{CB}\left ( D_{r}\left ( B_{D}^{1} \right ) \right ) \right ) \right ] \copyright  \left [ \texttt{RAAM}\left ( B_{E}^{3} \right ) \right ] \copyright \left [  f_{t}^{3\times 3} \left ( \texttt{RAAM}\left ( B_{E}^{4} \right ) \right )\right ]
    \label{Eq:D2}
\end{equation}
\begin{equation}
    B_{D}^{3}= \left [ f_{t}^{3\times 3}\left ( \texttt{CB}\left ( D_{r}\left ( B_{D}^{2} \right ) \right ) \right ) \right ] \copyright  \left [ \texttt{RAAM}\left ( B_{E}^{2} \right ) \right ] \copyright \left [  f_{t}^{3\times 3} \left ( \texttt{RAAM}\left ( B_{E}^{3} \right ) \right )\right ]
    \label{Eq:D3}
\end{equation}
\begin{equation}
    B_{D}^{4}= \left [ f_{t}^{3\times 3}\left ( \texttt{CB}\left ( D_{r}\left ( B_{D}^{3} \right ) \right ) \right ) \right ] \copyright  \left [ \texttt{RAAM}\left ( B_{E}^{1} \right ) \right ] \copyright \left [  f_{t}^{3\times 3} \left ( \texttt{RAAM}\left ( B_{E}^{2} \right ) \right )\right ]
    \label{Eq:D4}
\end{equation}

The final predicted mask of the proposed EUIS-Net is computed in (Eq.\ref{Eq:F}). 

\begin{equation}
    P_{mask}=\sigma \left ( f^{1\times 1}\left ( B_{D}^{4} \right ) \right )
    \label{Eq:F}
\end{equation}

Where $f^{1\times 1}$ is the convolutional operation with a kernel size ($1\times 1$) and $\sigma$ is the sigmoid operation.

\subsection{Region Aware Attention Module}

The region-aware attention module \cite{naveed2024ra} employs a composite pooling strategy, which combines maximum and average pooling in a cascade to learn region-aware features, as shown in (Fig. \ref{fig:modules}-(a)). This is achieved by promoting region-wise important information during the forward pass and ensuring region-dependent flow during back-propagation. The proposed approach effectively discriminates foreground pixels from background pixels by maximising the value resulting from the combined use of average and max pooling features. The computational steps involved in RAAM are briefly described in this section. Let $x$ be the input tensor of the model. The intermediate outputs ($I_{1}, I_{2}$) of the RAAM module are computed as (Eqs. \ref{eq:RAAM1}--\ref{eq:RAAM2})

\begin{equation}
    I_{1}=M_{p}\left ( \texttt{GAP}\left ( \Re\left ( B_{n}\left ( f^{3\times 3}\left ( x \right ) \right ) \right ) \right ) \right )
    \label{eq:RAAM1}
\end{equation}

\begin{equation}
    I_{2}=\texttt{GAP}\left ( M_{p}\left ( \Re\left ( B_{n}\left ( f^{3\times 3}\left ( x \right ) \right ) \right ) \right ) \right )
    \label{eq:RAAM2}
\end{equation}
where $\texttt{GAP}$ is the global average pooling. The final output ($y$) of the RAAM module is then computed by Eq. \ref{eq:RAAM3}.

\begin{equation}
y=\texttt{CCA}\left ( \Re\left ( B_{n}\left ( f^{3\times 3}\left ( x \right ) \right ) \right )
 \right )\times I_{1} \times I_{2} \times x    \label{eq:RAAM3}
\end{equation}

where $\texttt{CCA}$ is the cross channel averaging operation.

\subsection{Channel and Spatial Attention Module}

In our network, spatial and channel attention mechanisms are integrated within a residual block (Fig. \ref{fig:modules}-(b)). The interstitial association of features is employed to generate the spatial attention map, which focusses on the 'where' aspect of the input data. In contrast, channel attention serves as an informative component that complements spatial attention by focussing on the 'what' aspect. This combination enhances the ability of the network to selectively emphasise significant features both spatially and across channels. The output of the channel attention output is denoted by $(C_{A})$ and is computed as given in (Eq. \ref{Eq4}).

\begin{equation}
    C_{A}=E_{3}\otimes f^{3\times 3}\left ( \Re\left ( f^{3\times 3}\left [ \texttt{GAP} \left ( E_{3} \right ) \oplus \texttt{GMP} \left ( E_{3} \right ) \right ] \right ) \right )
    \label{Eq4}
\end{equation}
Where $\otimes$ is point-wise multiplication and $\oplus$ is the addition operator. $\texttt{GAP}$ and $\texttt{GMP}$ are the global average pooling and global max pooling operations, respectively.

The spatial attention output is denoted by $(S_{A})$ and computed as given in (Eq. \ref{Eq5}).

\begin{equation}
S_{A}=C_{A}\otimes \Re \left ( f^{3\times 3}\left ( C_{p}\left ( C_{A} \right ) \right ) \right )
    \label{Eq5}
\end{equation}
Where $C_{p}$ is the channel pooling operation. Once the feature information is extracted and refined from the encoder blocks and attention mechanisms this information is fed to the decoder blocks to reconstruct the feature maps and restore the spatial dimensions to the given input.
\section{Results and Discussion}\label{experimentalResults}

\subsection{Datasets}

The proposed EUIS-Net model was evaluated in two challenging benchmark ultrasound imaging datasets (Table \ref{tab:datasets}), namely BUSI \cite{BUSIdataset} for the segmentation of breast cancer and DDTI \cite{DDTI} for the segmentation of thyroid nodules. Both datasets are publicly available and provide GT masks for the evaluation of image segmentation methods. Performance evaluation was performed using a five-fold cross-validation method due to the unavailability of a separate test set.

\subsection{Evaluation Criteria}
The segmentation performance of EUIS-Net was evaluated and compared with SOTA methods using several metrics, including the Jaccard index ($J$, equal to IoU), Dice similarity coefficient ($D$), accuracy (A$_{cc}$), sensitivity (S$_n$), and specificity (S$_p$). These metrics were calculated according to their definitions:

\begin{equation}
J =\frac{T_{P}}{T_{P}+F_{P}+F_{N}},
\label{eq:Jindex}
\end{equation}

\begin{equation}
D =\frac{2 \times T_{P}}{2\times T_{P}+F_{P}+F_{N}},
\label{eq:Dice}
\end{equation}

\begin{equation}
A_{cc}=\frac{T_{P}+T_{N}}{T_{P}+T_{N}+F_{P}+F_{N}},
\label{eq:acc}
\end{equation}

\begin{equation}
S_{n}=\frac{T_{P}}{T_{P}+F_{N}},
\label{eq:sn}
\end{equation}

\begin{equation}
S_{p}=\frac{T_{N}}{T_{N}+F_{P}},
\label{eq:sp}
\end{equation}

\noindent where $T_{P}$, $T_{N}$, $F_{P}$, and $F_{N}$ denote the number of true positives, true negatives, false positives, and false negatives, respectively.

\begin{table*}[h]
\centering
\resizebox{1.0\textwidth}{!}{%
\caption{Details of the ultrasound image datasets used for the proposed EUIS-Net evaluation.}
\label{tab:datasets}
\begin{tabular}{lllllllll}
\hline
\textbf{Dataset} & \textbf{Reference} & \textbf{Image Format} & \textbf{Original Resolution} & \textbf{Image Count} & \textbf{Resized to} & \textbf{Dataset Split} & \textbf{Classes} & \textbf{Task / Additional Info} \\ \hline
BUSI & \cite{BUSIdataset} & .png & $\sim500\times500$ & 780 & $256\times256$ & 80\%:10\%:10\% & 3 & Breast ultrasound segmentation, age 25-75 \\ 
DDTI & \cite{BUSIdataset} & .png & $560\times360$, $280\times360$, $245\times360$ & 637 & $256\times256$ & 80\%:10\%:10\% & 2 & Thyroid nodule segmentation \\ 
\hline
\end{tabular}
}

\end{table*}

\begin{table}[!t]
  \centering
  \caption{Performance (mean ± std) comparison of EUIS-Net model with various SOTA methods on the breast lesion segmentation dataset BUSI \cite{BUSIdataset}.}
    \adjustbox{max width=\textwidth}{%
    \begin{tabular}{lccccc}
    \toprule
    \multirow{2}[4]{*}{\textbf{Method}} & \multicolumn{5}{c}{\textbf{Performance (\%)}} \\
    \cmidrule{2-6} & $J$ & $D$  & $A_{cc}$   & $S_{n}$   & $S{p}$ \\
    \midrule
    U-Net \cite{ronneberger2015u} & 67.77$\pm$2.35 & 76.96$\pm$3.67 & 95.48$\pm$3.60 & 78.33$\pm$4.35 & 96.13$\pm$2.07 \\
    % FPN \cite{lin2017feature} & 74.09 & 82.67 & -     & 85.39 & - \\
    % DeeplabV3+ \cite{chen2017rethinking} & 73.48 & 82.68 & -     & 83.37 & - \\
    % ConvEDNet \cite{lei2018segmentation} & 73.57 & 82.70 & -     & 85.51 & - \\
    UNet++ \cite{zhou2018unet++} & 76.85$\pm$3.13 & 76.22$\pm$3.59 & 97.97$\pm$1.93 & 78.61$\pm$3.27 & 98.86$\pm$2.23 \\
    ARU-GD \cite{maji2022attention} & 77.07$\pm$2.96 & 83.64$\pm$2.53 & 97.94$\pm$1.32 & 83.80$\pm$1.87 & 98.78$\pm$2.59 \\
    BCDU-Net \cite{azad2019bi} & 74.49$\pm$3.65 & 66.75$\pm$2.31 & 94.82$\pm$1.28 & 86.85$\pm$3.95 & 95.57$\pm$2.72 \\
    G-Net Light \cite{iqbal2022g} & 75.97$\pm$3.81 & 83.97$\pm$2.20 & 96.22$\pm$2.85   & 83.45$\pm$2.28 & 97.11$\pm$3.80\\
    MLR-Net \cite{iqbal2023mlr} & 76.09$\pm$2.04 & 83.67$\pm$3.06 & 96.65$\pm$2.55     & 84.39$\pm$1.13 & 98.04$\pm$1.44\\
    Swin-UNet \cite{cao2023swin} & 77.16$\pm$2.85 & 84.45$\pm$2.54 & 97.55$\pm$2.77 & 84.81$\pm$3.99 & 98.34$\pm$2.50 \\
    RA-Net \cite{naveed2024ra} & 77.48$\pm$2.08 & 84.68$\pm$2.91 & 97.82$\pm$1.92   & 85.37$\pm$3.32 & 98.14$\pm$1.95 \\
    \midrule
    \textbf{Proposed EUIS-Net} & \textbf{78.12$\pm$1.85} & \textbf{85.42$\pm$2.13} & \textbf{98.04$\pm$1.10} & \textbf{87.93$\pm$1.02} & \textbf{98.72$\pm$1.26} \\
    \bottomrule
    \end{tabular}%
  }
  \label{tab:BUSI}
\end{table}
\subsection{Experimental Setup}

For model training, the images (Table \ref{tab:datasets}) were augmented using contrast adjustments (with factors of [$\times 0.9, \times 1.1$]) and flipping operations (both in horizontal and vertical directions), which increased the size of the datasets by a factor of 5. The Adam optimiser was used with a maximum of 60 iterations and an initial learning rate of 0.001. In the absence of performance improvement on the validation set after five epochs, the learning rate was reduced by a quarter. To stop overfitting, an early stop strategy was implemented. The models were implemented through Keras using TensorFlow as the back-end and trained on an NVIDIA GTX 3090 GPU.

\begin{figure*}[h]
                 \centering
            \resizebox{1.0\textwidth}{!}{%
            \begin{tabular}{cccccccccccccccccccc}
            \\
               \hspace{5mm}Image  &&    \hspace{15mm} GT &&      \hspace{15mm} EUIS-Net &&  \hspace{3mm} Swin-UNet \cite{cao2023swin} &\hspace{2mm} BCDU-Net \cite{azad2019bi}&  \hspace{2mm} ARU-GD \cite{maji2022attention}&& MLR-Net \cite{iqbal2023mlr} &&  G-Net Light \cite{iqbal2022g} &&  UNet++ \cite{zhou2018unet++} &&  &  U-Net \cite{ronneberger2015u}&  \\ 
               \\
     
        \end{tabular}
    }
    \centering
     \resizebox{1.0\textwidth}{!}{%
     \begin{tabular}{cccccccccccccccccccc}
        \includegraphics[width=\textwidth]{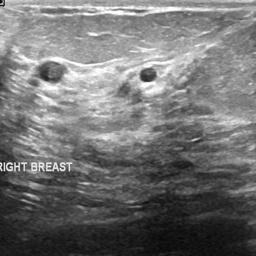} && \includegraphics[width=\textwidth]{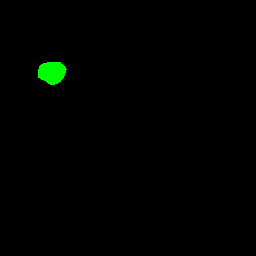} && \includegraphics[width=\textwidth]{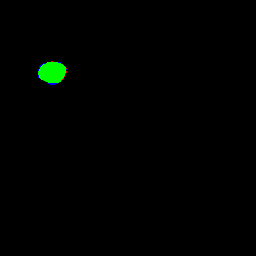} && \includegraphics[width=\textwidth]{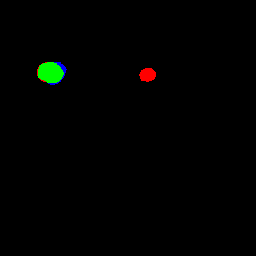} && \includegraphics[width=\textwidth]{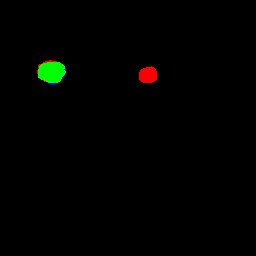} && \includegraphics[width=\textwidth]{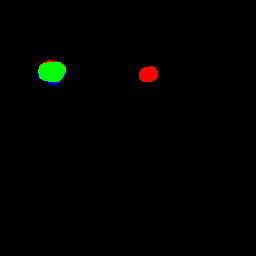} && \includegraphics[width=\textwidth]{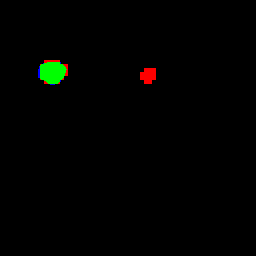} && \includegraphics[width=\textwidth]{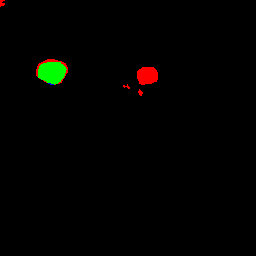} && \includegraphics[width=\textwidth]{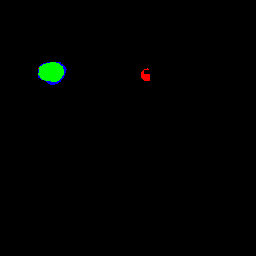} && \includegraphics[width=\textwidth]{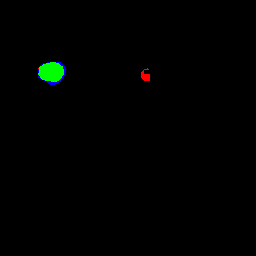} \\

\\
                \includegraphics[width=\textwidth]{Visual_Results/BUSI/4_im.png} && \includegraphics[width=\textwidth]{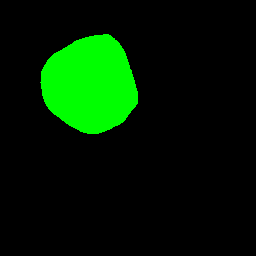} && \includegraphics[width=\textwidth]{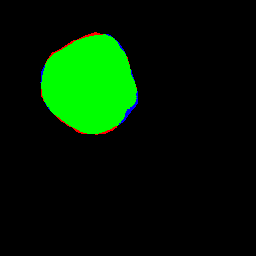} && \includegraphics[width=\textwidth]{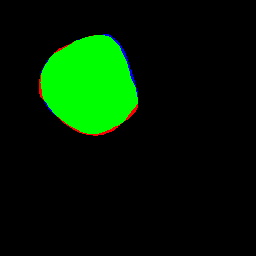} && \includegraphics[width=\textwidth]{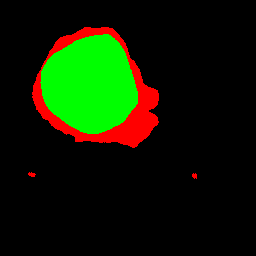} && \includegraphics[width=\textwidth]{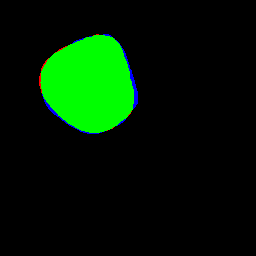} && \includegraphics[width=\textwidth]{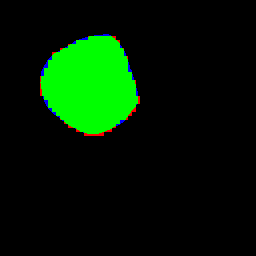} && \includegraphics[width=\textwidth]{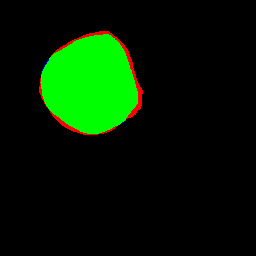} && \includegraphics[width=\textwidth]{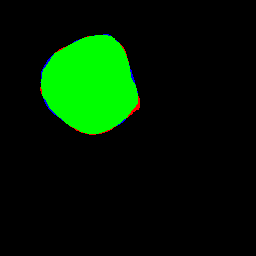} && \includegraphics[width=\textwidth]{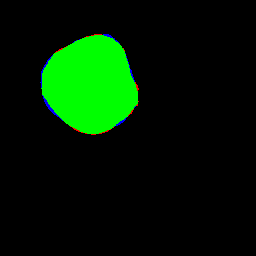} \\
                \\
                
                \includegraphics[width=\textwidth]{Visual_Results/BUSI/5_im.png} && \includegraphics[width=\textwidth]{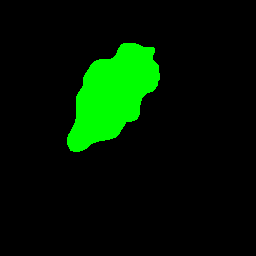} && \includegraphics[width=\textwidth]{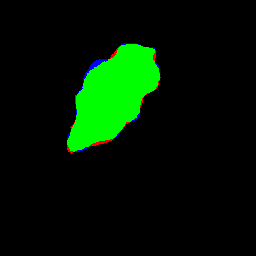} && \includegraphics[width=\textwidth]{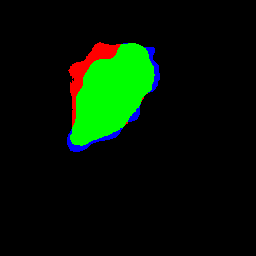} && \includegraphics[width=\textwidth]{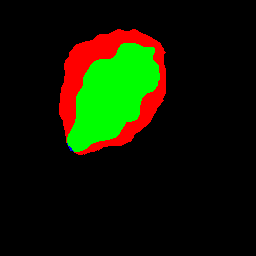} && \includegraphics[width=\textwidth]{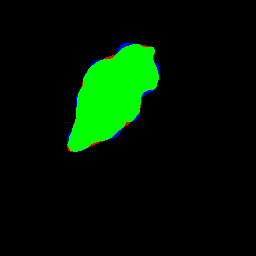} && \includegraphics[width=\textwidth]{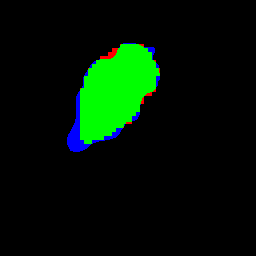} && \includegraphics[width=\textwidth]{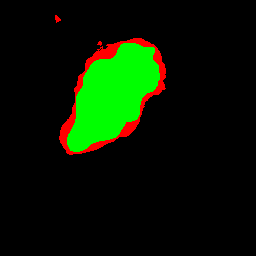} && \includegraphics[width=\textwidth]{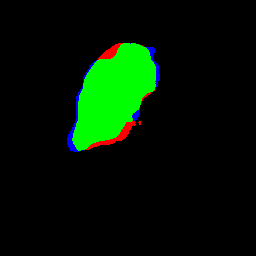} && \includegraphics[width=\textwidth]{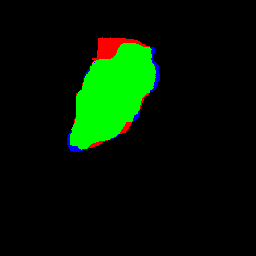} \\

                \\
        \end{tabular}
    }
    \caption{Visual performance comparison of the proposed EUIS-Net on BUSI \cite{BUSIdataset} dataset.}
    \label{fig:Vis_BUSI}
\end{figure*}

\begin{figure*}[h]

                 \centering
            \resizebox{1.0\textwidth}{!}{%
            \begin{tabular}{cccccccccccccccccccc}
            \\
\hspace{5mm}Image  &&    \hspace{15mm} GT &&      \hspace{15mm} EUIS-Net &&  \hspace{3mm} Swin-UNet \cite{cao2023swin} &\hspace{2mm} BCDU-Net \cite{azad2019bi}&  \hspace{2mm} ARU-GD \cite{maji2022attention}&& MLR-Net \cite{iqbal2023mlr} &&  G-Net Light \cite{iqbal2022g} &&  UNet++ \cite{zhou2018unet++} &&  &  U-Net \cite{ronneberger2015u}&  \\ 
               \\

        \end{tabular}
    }

    \centering
     \resizebox{1.0\textwidth}{!}{%
     
     \begin{tabular}{cccccccccccccccccccc}
        \includegraphics[width=\textwidth]{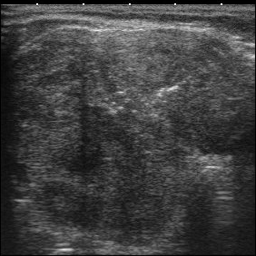} && \includegraphics[width=\textwidth]{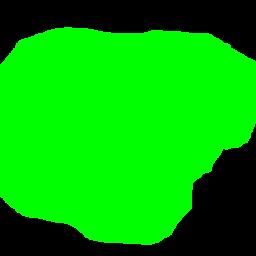} && \includegraphics[width=\textwidth]{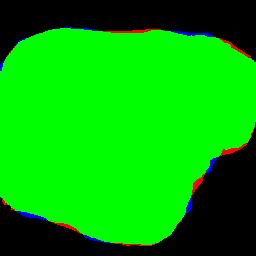} && \includegraphics[width=\textwidth]{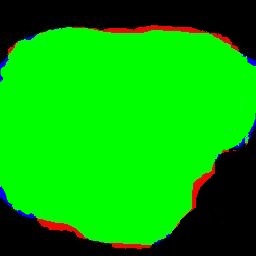} && \includegraphics[width=\textwidth]{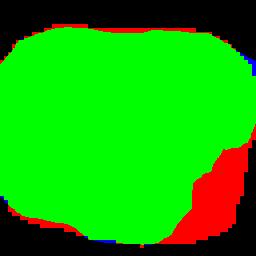} && \includegraphics[width=\textwidth]{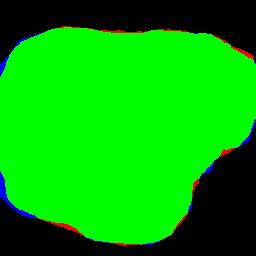} && \includegraphics[width=\textwidth]{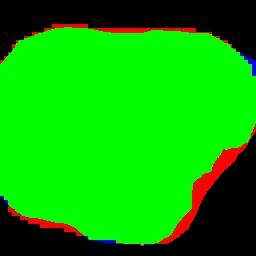} && \includegraphics[width=\textwidth]{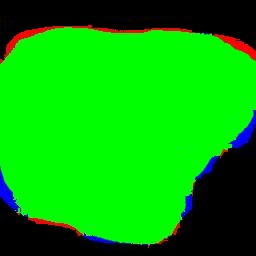} && \includegraphics[width=\textwidth]{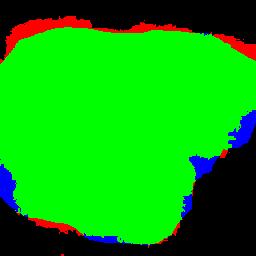} && \includegraphics[width=\textwidth]{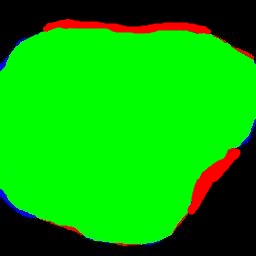} \\

\\
        \includegraphics[width=\textwidth]{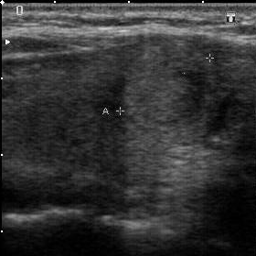} && \includegraphics[width=\textwidth]{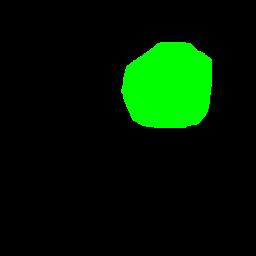} && \includegraphics[width=\textwidth]{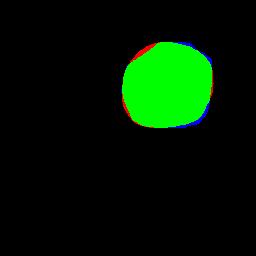} && \includegraphics[width=\textwidth]{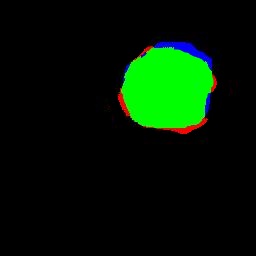} && \includegraphics[width=\textwidth]{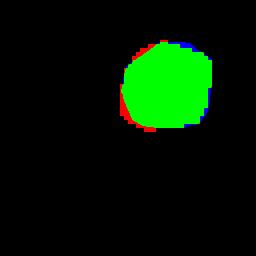} && \includegraphics[width=\textwidth]{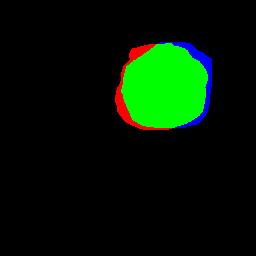} && \includegraphics[width=\textwidth]{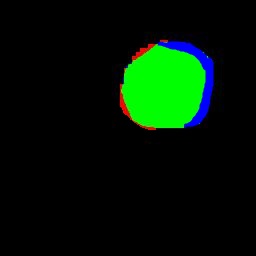} && \includegraphics[width=\textwidth]{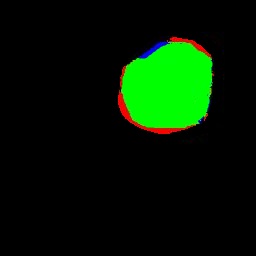} && \includegraphics[width=\textwidth]{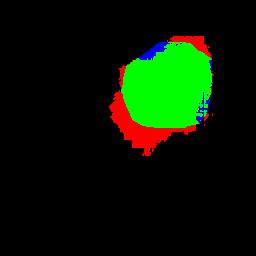} && \includegraphics[width=\textwidth]{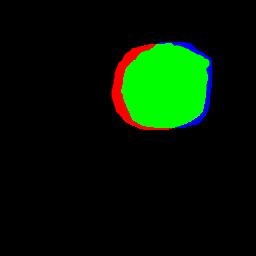} \\

\\
        \includegraphics[width=\textwidth]{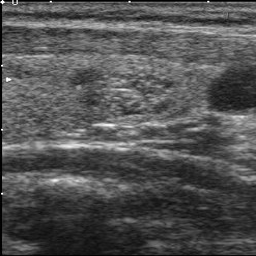} && \includegraphics[width=\textwidth]{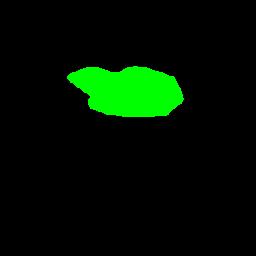} && \includegraphics[width=\textwidth]{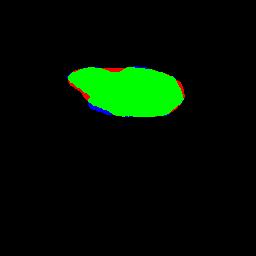} && \includegraphics[width=\textwidth]{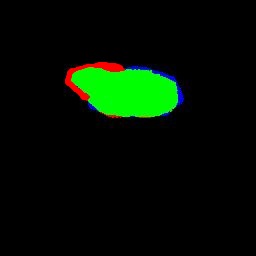} && \includegraphics[width=\textwidth]{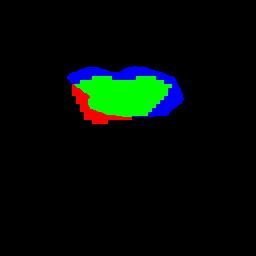} && \includegraphics[width=\textwidth]{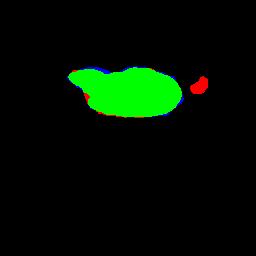} && \includegraphics[width=\textwidth]{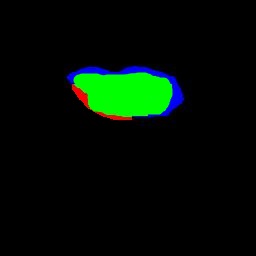} && \includegraphics[width=\textwidth]{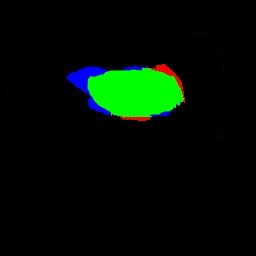} && \includegraphics[width=\textwidth]{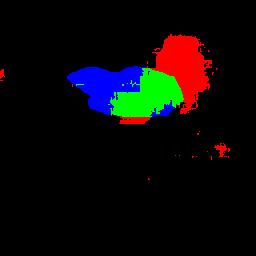} && \includegraphics[width=\textwidth]{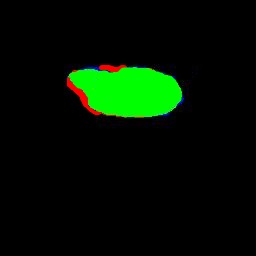} \\

        \end{tabular}
    }

    \caption{Visual performance comparison of the proposed EUIS-Net on DDTI \cite{DDTI} dataset.}
    \label{fig:Vis_DDTI}
\end{figure*}

\begin{table}[!t]
    \centering
    \caption{Performance (mean ± std) comparison of EUIS-Net with various SOTA methods on the thyroid nodule segmentation dataset DDTI \cite{DDTI}.}
    \adjustbox{max width=\textwidth}{%
    \begin{tabular}{lccccc}
    \toprule
    \multirow{2}[4]{*}{\textbf{Method}} & \multicolumn{5}{c}{\textbf{Performance (\%)}} \\
    \cmidrule{2-6}          & $J$ & $D$  & $A_{cc}$   & $S_{n}$   & $S_{p}$ \\
    \midrule
    U-Net \cite{ronneberger2015u}  & 74.76$\pm$1.36 & 84.08$\pm$3.19 & 96.55$\pm$2.48 & 85.50$\pm$3.09 & 97.57$\pm$1.61 \\
    UNet++ \cite{zhou2018unet++} & 74.76$\pm$3.46 & 84.08$\pm$2.27 & 96.55$\pm$2.51 & 85.50$\pm$3.39 & 97.57$\pm$1.34 \\
    BCDU-Net \cite{azad2019bi} & 77.79$\pm$1.90 & 79.49$\pm$2.27 & 93.22$\pm$2.51 & 82.31$\pm$3.39 & 94.34$\pm$1.34 \\
    ARU-GD \cite{maji2022attention} & 77.07$\pm$1.90 & 83.64$\pm$2.56 & 97.94$\pm$2.55 & 83.80$\pm$4.20 & 98.78$\pm$2.35 \\
    G-Net Light \cite{iqbal2022g} & 80.76$\pm$2.42 & 85.59$\pm$1.80 & 97.79$\pm$2.65 & 85.23$\pm$3.74 & 98.98$\pm$2.01  \\
    MLR-Net \cite{iqbal2023mlr} & 82.66$\pm$2.14 & 85.72$\pm$1.02 & 97.91$\pm$2.60 & 79.54$\pm$4.28 & 98.82$\pm$2.18  \\
    Swin U-Net \cite{cao2023swin} & 83.44$\pm$2.49 & 86.86$\pm$2.45 & 96.93$\pm$2.18 & 86.42$\pm$2.39 & 97.98$\pm$2.05\\
    RA-Net \cite{naveed2024ra} & 83.43$\pm$1.87 & 86.01$\pm$1.78 & 97.98$\pm$2.30 & 82.21$\pm$3.07 & 98.88$\pm$1.13  \\   
    \midrule
    \textbf{Proposed EUIS-Net} & \textbf{84.73$\pm$1.19} & \textbf{89.01$\pm$1.01} & \textbf{98.15$\pm$1.24} & \textbf{88.13$\pm$1.18} & \textbf{99.05$\pm$1.03} \\
    \bottomrule
    \end{tabular}%
    }
    \label{tab:DDTI}
\end{table}

\begin{figure*}[h]
                 \centering
            \resizebox{1.0\textwidth}{!}{%
            \begin{tabular}{cccccccccccccccccccc}
            \\
\hspace{5mm}Image  &&    \hspace{15mm} GT &&      \hspace{15mm} EUIS-Net &&  \hspace{3mm} Swin-UNet \cite{cao2023swin} &\hspace{2mm} BCDU-Net \cite{azad2019bi}&  \hspace{2mm} ARU-GD \cite{maji2022attention}&& MLR-Net \cite{iqbal2023mlr} &&  G-Net Light \cite{iqbal2022g} &&  UNet++ \cite{zhou2018unet++} &&  &  U-Net \cite{ronneberger2015u}&  \\ 
               \\

        \end{tabular}
    }
    \centering
     \resizebox{1.0\textwidth}{!}{%
     
     \begin{tabular}{cccccccccccccccccccc}
        \includegraphics[width=\textwidth]{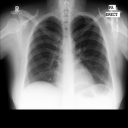} && \includegraphics[width=\textwidth]{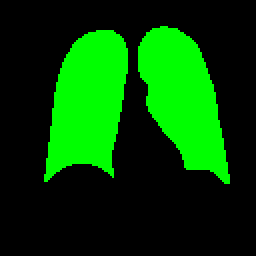} && \includegraphics[width=\textwidth]{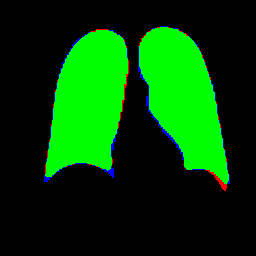} && \includegraphics[width=\textwidth]{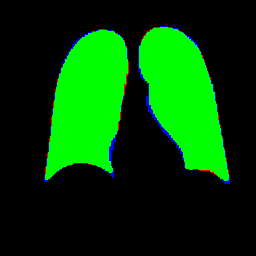} && \includegraphics[width=\textwidth]{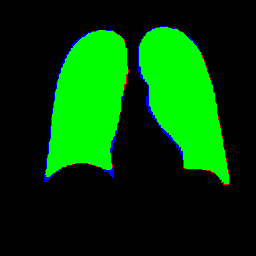} && \includegraphics[width=\textwidth]{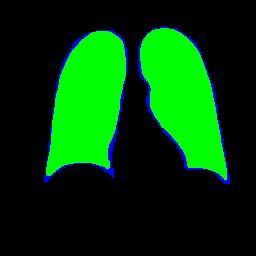} && \includegraphics[width=\textwidth]{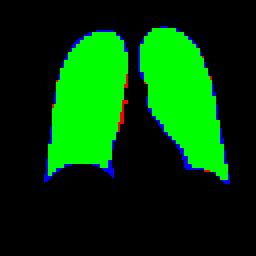} && \includegraphics[width=\textwidth]{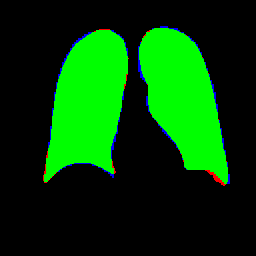} && \includegraphics[width=\textwidth]{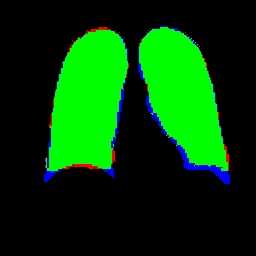} && \includegraphics[width=\textwidth]{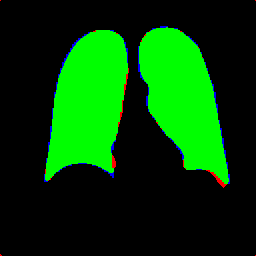} \\

\\

       \includegraphics[width=\textwidth]{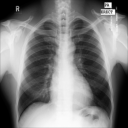} && \includegraphics[width=\textwidth]{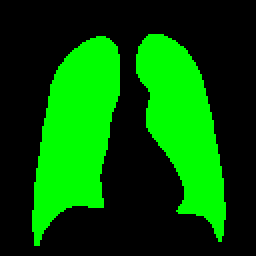} && \includegraphics[width=\textwidth]{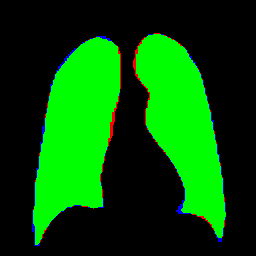} && \includegraphics[width=\textwidth]{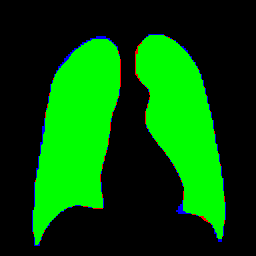} && \includegraphics[width=\textwidth]{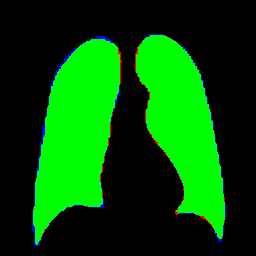} && \includegraphics[width=\textwidth]{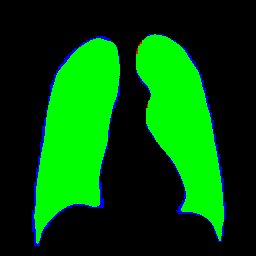} && \includegraphics[width=\textwidth]{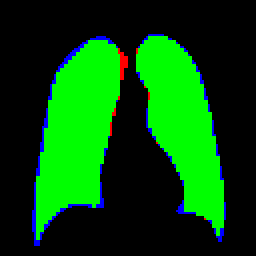} && \includegraphics[width=\textwidth]{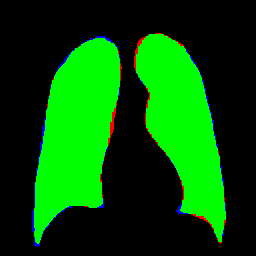} && \includegraphics[width=\textwidth]{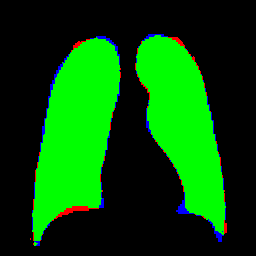} && \includegraphics[width=\textwidth]{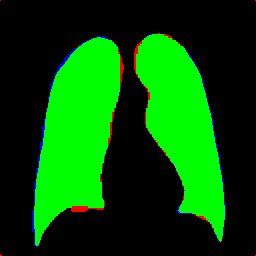} \\

\\
       \includegraphics[width=\textwidth]{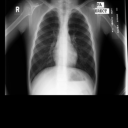} && \includegraphics[width=\textwidth]{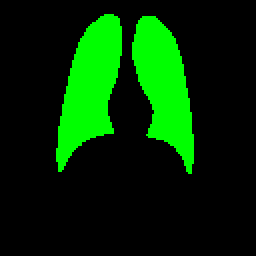} && \includegraphics[width=\textwidth]{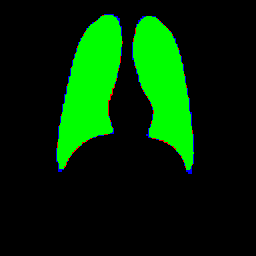} && \includegraphics[width=\textwidth]{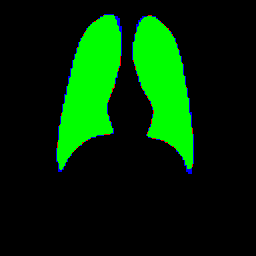} && \includegraphics[width=\textwidth]{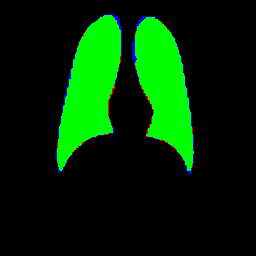} && \includegraphics[width=\textwidth]{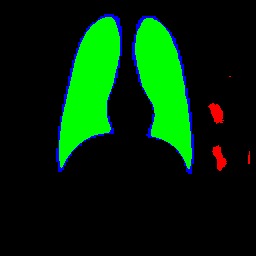} && \includegraphics[width=\textwidth]{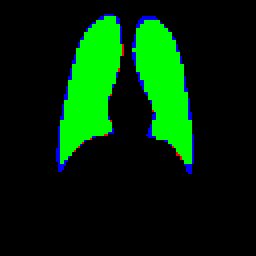} && \includegraphics[width=\textwidth]{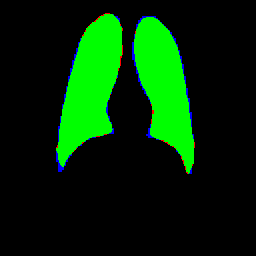} && \includegraphics[width=\textwidth]{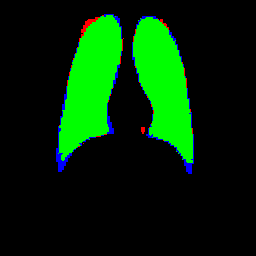} && \includegraphics[width=\textwidth]{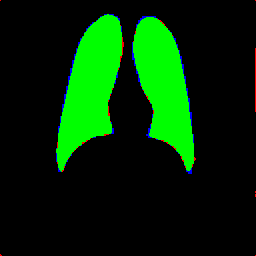} \\

        \end{tabular}
    }
         
    \caption{Visual performance comparison of the proposed EUIS-Net on MC \cite{MC_Dataset} dataset.}
    \label{fig:Vis_MC}
\end{figure*}

\begin{table}[h]
    \centering
    \caption{Performance (mean ± std) comparison of the proposed EUIS-Net with various SOTA methods for chest X-ray image segmentation on the MC \cite{MC_Dataset} dataset.}
    \adjustbox {max width=\textwidth}{%
    \begin{tabular}{lccccc}
    \toprule
    \multirow{2}[4]{*}{\textbf{Method}} & \multicolumn{5}{c}{\textbf{Performance (\%)}} \\
    \cmidrule{2-6}          & $J$ & $D$  & $A_{cc}$   & $S_{n}$   & $S_{p}$ \\
    \midrule
    U-Net \cite{ronneberger2015u}  & 94.38$\pm$2.10 & 97.10$\pm$1.58 & 98.53$\pm$1.18 & 95.99$\pm$2.04 & 99.40$\pm$0.44 \\
    UNet++ \cite{zhou2018unet++} & 94.57$\pm$2.82 & 97.20$\pm$1.82 & 98.59$\pm$1.20 & 96.28$\pm$2.03 & 99.36$\pm$0.62 \\
    ARU-GD \cite{maji2022attention} & 94.65$\pm$2.35 & 97.25$\pm$1.48 & 98.62$\pm$1.30 & 95.97$\pm$1.77 & \textbf{99.50$\pm$0.48} \\
    BCDU-Net \cite{azad2019bi} & 94.53$\pm$2.59 & 97.19$\pm$1.57 & 98.57$\pm$1.20 & 96.00$\pm$1.11 & 99.43$\pm$0.32 \\
    G-Net Light \cite{iqbal2022g} & 94.47$\pm$2.84& 97.22$\pm$1.31 & 98.03$\pm$1.69 & 95.09$\pm$1.55 & 99.42$\pm$0.48 \\
    MLR-Net \cite{iqbal2023mlr} & 94.80$\pm$2.29 & 97.34$\pm$1.85 & 98.70$\pm$0.70 & 95.57$\pm$1.68 & 99.02$\pm$0.85 \\
    Swin-UNet \cite{cao2023swin}& 90.69$\pm$1.40 & 95.07$\pm$1.62 & 97.61$\pm$0.65 & 94.34$\pm$1.39 & 98.65$\pm$0.59 \\
    RA-Net \cite{naveed2024ra} & 94.88$\pm$1.80 & 97.49$\pm$1.28 & 98.87$\pm$0.21 & 96.89$\pm$1.67 & 99.20$\pm$0.40 \\ 
    \hline
    \textbf{Proposed EUIS-Net} & \textbf{96.49$\pm$1.59} & \textbf{98.21$\pm$1.09} & \textbf{99.14$\pm$0.14} & \textbf{98.40$\pm$1.05} & \textbf{99.33$\pm$0.11} \\   
    \bottomrule
    \end{tabular}%
    }
  \label{tab:MC}
\end{table}
\subsection{Comparisons with SOTA Methods}

For both data sets (Table~\ref{tab:datasets}) we compared EUIS-Net with eight state-of-the-art medical image segmentation methods, including ARU-GD \cite{maji2022attention}, BCDU-Net \cite{azad2019bi}, G-Net Light \cite{iqbal2022g}, MLR-Net \cite{iqbal2023mlr}, RA-Net \cite{naveed2024ra}, Swin-Unet \cite{cao2023swin}, U-Net \cite{ronneberger2015u} and UNet++ \cite{zhou2018unet++}.

Table \ref{tab:BUSI} presents the statistical comparison of the proposed EUIS-Net with the other methods. Compared to other methods, the Jaccard index ($J$) of the proposed EUIS-Net is improved by 0.64\%, .96\%, 2.03\%, 2.15\%, 3.63\%, 1.05\%, 1.27\%, and 10.35\% than RA-Net \cite{naveed2024ra}, Swin-Unet \cite{cao2023swin}, MLR-Net \cite{iqbal2023mlr}, G-Net Light \cite{iqbal2022g}, BCDU-Net \cite{azad2019bi}, ARU-GD \cite{maji2022attention}, UNet++ \cite{zhou2018unet++}, and U-Net \cite{ronneberger2015u}, respectively on the BUSI dataset \cite{BUSIdataset}. Figure \ref{fig:Vis_BUSI} presents the visual results of different challenges in the segmentation of breast cancer. The proposed EUIS-Net achieved the best segmentation results, which closely resemble GT data, even for breast cancer images with varying sizes and irregular shapes.
Table \ref{tab:DDTI} presents the statistical comparison of the proposed EUIS-Net with the other methods. Compared to other methods, the Jaccard index ($J$) of the proposed EUIS-Net is improved by 1.3\%, 1.29\%, 2.07\%, 3.97\%, 7.66\%, 6.94\%, 9.97\%, and 10.67\% than RA-Net \cite{naveed2024ra}, Swin-Unet \cite{cao2023swin}, MLR-Net \cite{iqbal2023mlr}, G-Net Light \cite{iqbal2022g}, BCDU-Net \cite{azad2019bi}, ARU-GD \cite{maji2022attention}, UNet++ \cite{zhou2018unet++}, and U-Net \cite{ronneberger2015u}, respectively on the DDTI dataset \cite{DDTI} for thyroid nodule segmentation. Figure \ref{fig:Vis_DDTI} presents the visual results of different challenges in the segmentation of breast cancer. The proposed EUIS-Net achieved the best segmentation results, which closely resemble GT data, even for thyroid nodule images with varying sizes and irregular shapes.

The generalisation of the proposed EUIS-Net is evaluated on the MC \cite{MC_Dataset} data set for chest X-ray image segmentation. Table \ref{tab:MC} presents the statistical comparison of the proposed EUIS-Net with the other methods. Compared to other methods, the Jaccard index ($J$) of the proposed EUIS-Net is improved by 1.3\%, 1.29\%, 2.07\%, 3.97\%, 7.66\%, 6.94\%, 9.97\%, and 10.67\% than RA-Net \cite{naveed2024ra}, Swin-Unet \cite{cao2023swin}, MLR-Net \cite{iqbal2023mlr}, G-Net Light \cite{iqbal2022g}, BCDU-Net \cite{azad2019bi}, ARU-GD \cite{maji2022attention}, UNet++ \cite{zhou2018unet++}, and U-Net \cite{ronneberger2015u}, respectively on the MC dataset \cite{MC_Dataset} for chest X-ray segmentation. Figure \ref{fig:Vis_MC} presents the visual results of different chest X-ray image segmentation challenges. The proposed EUIS-Net achieved the best segmentation results, which closely resemble GT data.

% \begin{table}[!htbp]
%   \centering
%   \caption{Computational requirements comparisons of the EUIS-Net with other methods.}
%   \resizebox{1.0\textwidth}{!}{%

%     \begin{tabular}{lcc}
%     \toprule
%     \multicolumn{1}{l}{Method} & Params (M) & J in (\%)\\
%     \midrule
%     U-Net \cite{ronneberger2015u} & 9.2 & 67.77\\
%     UNet++ \cite{zhou2018unet++} & 9.1  & 76.85\\
%     ARU-GD \cite{maji2022attention} & 33.5 & 77.07\\
%    BCDU-Net \cite{azad2019bi} & 9.5 & 74.49 \\
%     G-Net Light \cite{iqbal2022g} & 0.39 & 75.97\\
%     MLR-Net \cite{iqbal2023mlr} & 0.81  & 76.09\\
%     Swin-UNet \cite{cao2023swin} & 27.3 & 77.16\\
%    RA-Net \cite{naveed2024ra} & 7.51  & 77.48 \\
   
%     \hline
%     \textbf{EUIS-Net} & \textbf{1.8} & \textbf{78.12} \\
%     \bottomrule
%     \end{tabular}%
%     }

%   \label{tab:timingComparison}%
% \end{table}%

\section{Conclusions}
\label{conclusions}

We have developed EUIS-Net, a novel and effective methodology for segmenting ultrasound images with precision, aimed at overcoming challenges like inherent noise and unpredictability of ultrasound images. To address these challenges, we proposed EUIS-Net, a CNN network designed to segment ultrasound images efficiently and precisely. The proposed EUIS-Net uses four encoder-decoder blocks, resulting in a notable decrease in computational complexity while achieving excellent performance. The proposed EUIS-Net integrates both channel and spatial attention mechanisms into the bottleneck to improve feature representation and collect significant contextual information. In addition, EUIS-Net incorporates a region-aware attention module (RAAM) into the skip connections, which enhances the ability to concentrate on lesion regions. When evaluated on two publicly available benchmark ultrasound image datasets, EUIS-Net outperformed a variety of recent methods. Despite its exceptional performance, we have identified areas for further improvement. We suggest employing EUIS-Net in a semi-supervised manner to reduce data requirements for training. EUIS-Net is suitable not only for immediate medical imaging applications in ultrasound image segmentation but also holds promise for adaptation and expansion to other medical imaging and segmentation tasks.

% \begin{credits}
% \subsubsection{\ackname} A bold run-in heading in small font size at the end of the paper is
% used for general acknowledgments, for example: This study was funded
% by X (grant number Y).

% \subsubsection{\discintname}
% It is now necessary to declare any competing interests or to specifically
% state that the authors have no competing interests. Please place the
% statement with a bold run-in heading in small font size beneath the
% (optional) acknowledgments\footnote{If EquinOCS, our proceedings submission
% system, is used, then the disclaimer can be provided directly in the system.},
% for example: The authors have no competing interests to declare that are
% relevant to the content of this article. Or: Author A has received research
% grants from Company W. Author B has received a speaker honorarium from
% Company X and owns stock in Company Y. Author C is a member of committee Z.
% \end{credits}
%
% ---- Bibliography ----
%
% BibTeX users should specify bibliography style 'splncs04'.
% References will then be sorted and formatted in the correct style.
%
% \bibliographystyle{splncs04}
% \bibliography{mybibliography}
%
\bibliographystyle{IEEEtran}
\bibliography{References}
% \begin{thebibliography}{8}

% \bibitem{ref_article1}
% Author, F.: Article title. Journal \textbf{2}(5), 99--110 (2016)

% \bibitem{ref_lncs1}
% Author, F., Author, S.: Title of a proceedings paper. In: Editor,
% F., Editor, S. (eds.) CONFERENCE 2016, LNCS, vol. 9999, pp. 1--13.
% Springer, Heidelberg (2016). \doi{10.10007/1234567890}

% \bibitem{ref_book1}
% Author, F., Author, S., Author, T.: Book title. 2nd edn. Publisher,
% Location (1999)

% \bibitem{ref_proc1}
% Author, A.-B.: Contribution title. In: 9th International Proceedings
% on Proceedings, pp. 1--2. Publisher, Location (2010)

% \bibitem{ref_url1}
% LNCS Homepage, \url{http://www.springer.com/lncs}, last accessed 2023/10/25
% \end{thebibliography}
\end{document}